\documentclass[prl,aps,twocolumn, showpacs]{revtex4}

\usepackage{capt-of}
\usepackage{graphicx}
\usepackage{amsmath}
\usepackage{bm}
\usepackage[latin1]{inputenc}
\usepackage{amstext}
\usepackage{amsxtra}
\newcommand{\bs}{\boldsymbol}
\newcommand{\beq}{\begin{equation}}
\newcommand{\eeq}{\end{equation}}
\newcommand{\bpm}{\begin{pmatrix}}
\newcommand{\epm}{\end{pmatrix}}
\newcommand{\bea}{\begin{eqnarray}}
\newcommand{\eea}{\end{eqnarray}}{}

\begin{document}

\title{Reaching Fractional Quantum Hall States with Optical Flux Lattices}
\date{14 December 2012}
\author{Nigel R. Cooper$^1$ and Jean Dalibard$^{2,3}$}
\affiliation{$^1$T.C.M. Group, Cavendish Laboratory, J.~J.~Thomson Avenue, Cambridge CB3~0HE, United Kingdom\\
 $^2$ Laboratoire Kastler Brossel, CNRS, UPMC, ENS, 24 rue Lhomond, F-75005 Paris, France\\
 $^3$ Collège de France, 11, place Marcelin Berthelot, 75005 Paris, France }



\begin{abstract}

  We present a robust scheme by which fractional quantum Hall states
  of bosons can be achieved for ultracold atomic gases.  We describe a new
  form of optical flux lattice, suitable for commonly used atomic
  species with groundstate angular momentum $J_g = 1$, for which the
  lowest energy band is topological and nearly dispersionless. Through
  exact diagonalization studies, we show that, even for moderate
  interactions, the many-body groundstates consist of bosonic
  fractional quantum Hall states, including the Laughlin state and the
  Moore-Read (Pfaffian) state.  Under realistic conditions, these phases are
  shown to have energy gaps that are larger than temperature
  scales achievable in ultracold gases.

\end{abstract}

\maketitle

There is intense interest in finding new settings in which topological phases of matter analogous to fractional quantum Hall (FQH) states appear. Ultracold atomic gases are ideal systems with which to achieve this goal: they allow studies of strong correlation phenomena for both fermions and bosons, and FQH physics can be approached for homogeneous fluids \cite{advances} as well as for atoms confined in optical lattices \cite{lewensteinrev}. 

 While existing theories of FQH-like phases in lattices have focussed on tight-binding models \cite{sorensen:086803,mollercooper-cf,mazzamr,PhysRevLett.106.236804,shenggusunsheng,regnaultlattice,hormozi},
one of the most promising routes to topological flat bands for ultracold atoms
is through optical flux lattices (OFLs) \cite{ofl,cooperdalibard,coopermoessner}. An OFL uses a set of
laser beams to produce a spatially periodic atom-laser coupling that
induces resonant transitions between two (or more) internal atomic
states.
The resulting energy bands, in particular the lowest one, have non-zero Chern numbers, and can be
made narrow in energy \cite{coopermoessner}. This opens the path to
experimental studies of novel strong correlation phenomena in topological
flat bands, notably the FQH effect of bosons. 

We present in this paper the first characterization of the many-body
ground state of bosons in an OFL. We start with the design of a novel
type of OFL, which fully exploits the structure of the most commonly
used (bosonic) atomic species. The scheme is robust since, by contrast
to some other OFL proposals \cite{ofl}, it does not require phase
locking between the various optical beams composing the lattice. For
optimized parameters its lowest band has Chern number 1 and is nearly
dispersionless, closely analogous to the lowest Landau level for
charged particles moving in a uniform magnetic field. We use exact
diagonalization to determine the many-body spectrum of a bosonic gas
in this OFL. We show that FQH ground states appear for relatively weak
atom interaction at the same filling factors as for a continuum Landau
level \cite{advances}. Our work provides a concrete experimental
scheme by which FQH states of bosons can be realized with large energy
scales.  Furthermore, it provides the first example of a non-Abelian
quantum Hall state (the $\nu=1$ Moore-Read state \cite{MooreR91}) in a
lattice model at high particle density with only two-body
interactions.

\begin{figure*}
\vskip-6mm
\hskip-5mm\includegraphics[width=4.2cm]{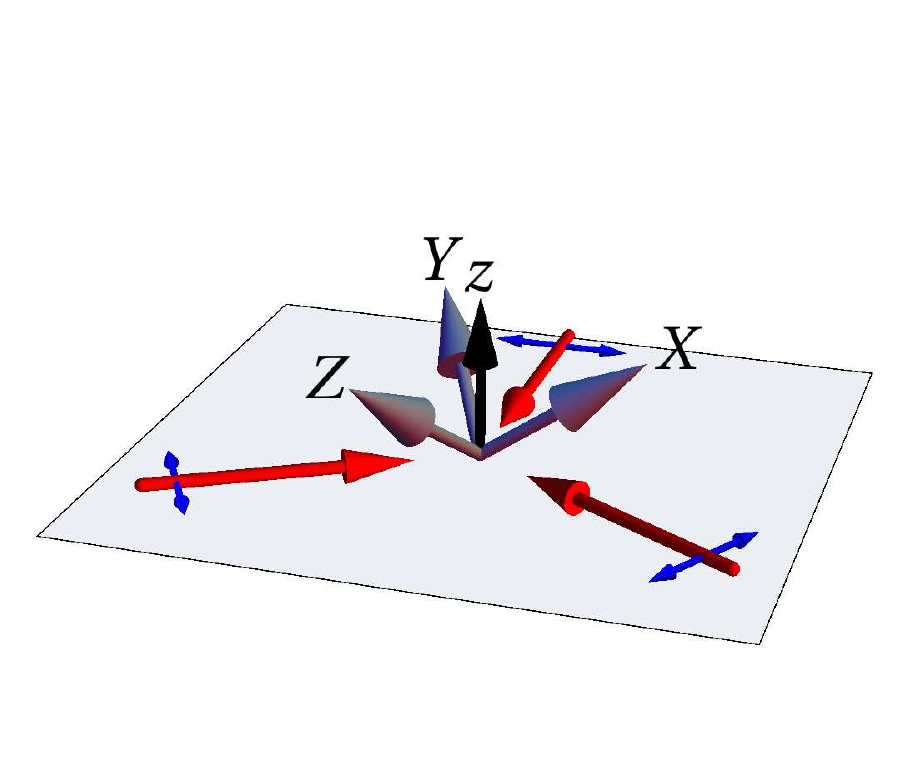}
\includegraphics[width=135mm]{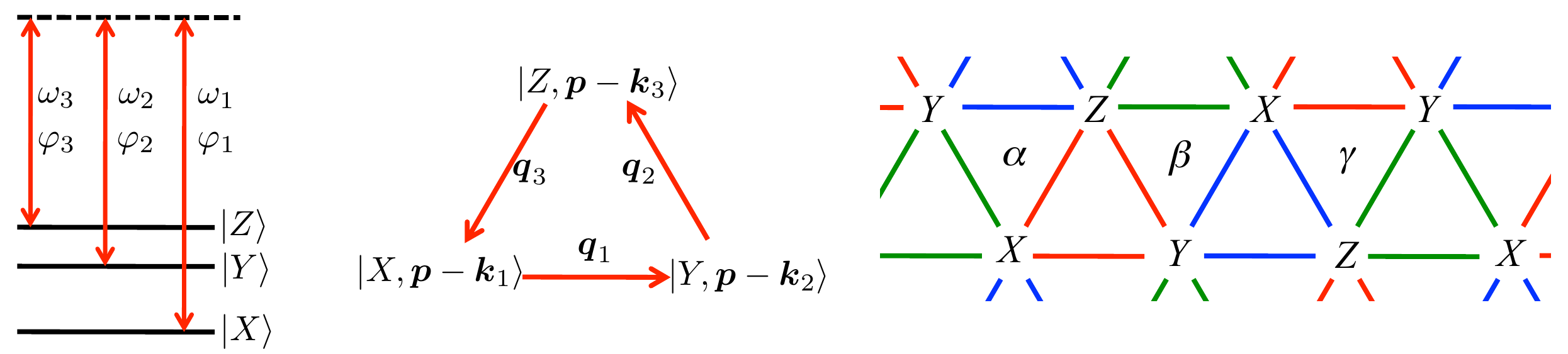}
\vskip-30mm
{(a) \hskip 32mm (b) \hskip 30mm (c) \hskip 35mm (d)  \hfill \hfill}
\vskip 25mm
\caption{(a) Atoms with a ground state with angular momentum $J_g=1$
are irradiated by three laser running waves propagating in the $xy$ plane, whose wave vectors $\bs k_i$, $i=1,2,3$ make an angle of $2\pi/3$ with each other. (b) Triplet of light frequencies $\omega_i$  ensuring that the three possible Raman transitions are resonantly driven.  (c) Graphic representation of three internal+momentum eigenstates, which are resonantly coupled by the laser beams whose frequencies are shown in (b). (d) Infinite array of internal+momentum eigenstates that are resonantly coupled when three triplets of frequencies $\omega_i$ (red), $\omega'_i$ (green), $\omega''_i$ (blue) are simultaneously applied (see also Table \ref{table}).
}
\label{fig:Fig1}
\end{figure*}

We focus in this paper on the case of atoms whose internal ground level has angular momentum $J_g=1$. This is the case for several stable bosonic isotopes of alkali metal species, namely $^7$Li, $^{23}$Na, $^{39}$K, $^{41}$K, $^{87}$Rb. We denote $|X\rangle$, $|Y\rangle$, $|Z\rangle$ a basis of this level, defined such that $\hat J_X|X\rangle=0$ (and similarly for $Y$ and $Z$). Here the set of directions $X,Y,Z$ represents an orthogonal trihedron of the physical space [see Figure \ref{fig:Fig1}(a)] and $\hat J_X$ stands for the component of the angular momentum operator along the $X$ direction. Note that one can replace $|X\rangle$, $|Y\rangle$, $|Z\rangle$ by  a triplet of internal states selected among a more complex level scheme, as proposed {\it e.g.} in~\cite{Campbell:2011}. Our scheme will apply as long as each pair of states can be coupled by a resonant two-photon Raman transition with a negligible spontaneous emission rate~\cite{external}. 

We assume that  $|X\rangle$, $|Y\rangle$, $|Z\rangle$ are the eigenstates of the atomic Hamiltonian in the absence of atom-laser coupling. We suppose that these three states are non-degenerate and non-equally spaced, and their energies are such that $E_X < E_Y<E_Z$, with $E_Z-E_Y \neq E_Y-E_X$.  For alkali atoms this situation can be reached by illuminating the atomic sample with microwaves close to the hyperfine resonance (see supplementary information). We denote by $z$ the $(1,1,1)$ direction of the $X,Y,Z$ trihedron, and assume that the centre-of-mass motion of the atoms along the direction $z$ is frozen. Therefore we consider in the following only the atomic motion in the perpendicular $xy$ plane [see Figure~\ref{fig:Fig1}(a)].

The atoms are irradiated with laser beams propagating in the $xy$ plane along three directions making an angle of $2\pi/3$ with each other. The three wave vectors are $\bs k_1=k/2\, (\sqrt 3\, \bs u_x +\bs u_y)$, $\bs k_2=k/2\, (-\sqrt 3\, \bs u_x +\bs u_y)$ and $\bs k_3=-k\bs u_y$, where $\{\bs u_x,\bs u_y\}$ is an orthogonal unit basis of the $xy$ plane. Here $k$ stands for the typical wave number of the laser beams~\cite{Note1}. We choose the frequency components in each laser beam so that an atom can undergo resonant Raman transitions between the three internal states, by absorbing  a photon in one wave and emitting a photon in a stimulated manner in another wave. The momentum change in such a transition is $\pm \bs q_i$, where $\bs q_i=\bs k_{i}-\bs k_{i+1}$. Here we set $\bs k_4\equiv \bs k_1$ and take $\hbar=1$ for simplicity. 

Suppose first that each beam $i$ consists only of a monochromatic plane wave with frequency $\omega_i$ and phase $\varphi_i$, and that the $\omega_i$'s are chosen such the three Raman conditions are fulfilled: $\omega_1-\omega_2=E_Y-E_X$, $\omega_2-\omega_3=E_Z-E_Y$ [and thus $\omega_1-\omega_3=E_Z-E_X$, see Figure \ref{fig:Fig1}(b)]. Each family of momentum eigenstates ${\cal F}(\bs p)=\{ |X,\bs p-\bs k_1\rangle, |Y,\bs p-\bs k_2\rangle, |Z,\bs p-\bs k_3\rangle \}$ generates a manifold that is globally stable with respect to atom-laser coupling. The three states of  ${\cal F}(\bs p)$ form an equilateral triangle in momentum space [Figure~\ref{fig:Fig1}(c)]. The coupling between the atom and the laser field can be written (see supplementary material)
\begin{eqnarray}
\hat{V}=- \Omega &\Big(& |Y\rangle \langle X| e^{i (\bs q_1\cdot \bs r +\varphi_1-\varphi_2)}+|Z\rangle \langle Y| e^{i (\bs q_2\cdot \bs r +\varphi_2-\varphi_3)}\nonumber \\
&+&|X\rangle \langle Z| e^{i (\bs q_3\cdot \bs r +\varphi_3-\varphi_1)}\Big)\ +\ \mbox{H.c.},
\label{eq:coupling}
\end{eqnarray}
where  H.c. stands for Hermitian conjugate. The amplitude and sign of the coupling strength $\Omega$ can be adjusted by tuning the intensity of the coupling lasers, and their detuning with respect to the atomic resonance. The fact that all three Raman transitions in Eq. (\ref{eq:coupling}) have the same amplitude is ensured by (i) taking the same intensity for each laser beam, (ii) choosing in-plane linear polarizations. A similar ring-coupling scheme has been used in~\cite{2012:Jimenez} to implement the Peierls substitution in a 1D optical lattice. However in~\cite{2012:Jimenez} only two laser Raman transitions were used and the ring was closed using radio frequency transitions, which is not appropriate for our purpose. 

With only one triplet of laser frequencies as in Fig.~\ref{fig:Fig1}(b), we do not produce the desired infinite periodic lattice for the atomic motion in momentum space \cite{coopermoessner}. However this goal can be reached by adding inside the beams $i$  two other triplets of frequency components $\omega'_i$ and $\omega''_i$, $i=1,2,3$. Here the roles are circularly exchanged with respect to the first triplet $\omega_i$: The $\omega'_i$ (resp.  $\omega''_i$) are such that  $\omega'_2-\omega'_3=E_Y-E_X$ and $\omega'_3-\omega'_1=E_Z-E_Y$ (resp. $\omega''_3-\omega''_1=E_Y-E_X$ and $\omega''_1-\omega''_2=E_Z-E_Y$).  In the following we suppose that the differences between the average frequencies $\bar \omega, \bar \omega', \bar \omega''$ of the triplets are much larger than the splittings $E_\alpha-E_\beta$. Processes involving the absorption of a photon from a frequency triplet and the stimulated emission of a photon in another triplet thus play a negligible role.

With the three frequency triplets acting simultaneously on an atom, the family of states that are coupled to a given initial state can be represented by the infinite lattice in momentum space shown in Fig.~\ref{fig:Fig1}(d). Since there are 3 possible Raman transitions and 3 possible pairs of beams to induce a given transition, the atom-laser coupling $\hat V$ generalizing (\ref{eq:coupling}) is now characterized by 9 matrix elements. These elements depend on the 9 phases $\varphi_i,\varphi'_i,\varphi''_i$ and are summarized in Table 1. From this Table, it is straightforward to write down explicitly the coupling $\hat V$. For example the three terms appearing in (\ref{eq:coupling}) correspond to the diagonal terms of the array of Table \ref{table}.

 \begin{table}[t]
\begin{center}
\begin{tabular}{c|c|c|c|}
 & $X \rightarrow Y$ & $Y \rightarrow Z$ & $Z \rightarrow X$ \\
 \hline
$\bs q_1$ & $e^{i(\varphi_1-\varphi_2)}$& $e^{i(\varphi''_1-\varphi''_2)}$    & $e^{i(\varphi'_1-\varphi'_2)}$ \\  
 \hline
$\bs q_2$ & $e^{i(\varphi'_2-\varphi'_3)}$& $e^{i(\varphi_2-\varphi_3)}$ & $e^{i(\varphi''_2-\varphi''_3)}$ \\
 \hline
$\bs q_3$ & $e^{i(\varphi''_3-\varphi''_1)}$& $e^{i(\varphi'_3-\varphi'_1)}$ & $e^{i(\varphi_3-\varphi_1)}$ \\
\hline
\end{tabular}
\end{center}
\caption{Phases of the Raman coupling matrix elements. 
Each line corresponds to a given momentum kick $\bs q_i=\bs k_i-\bs k_{i+1}$, and each column to a given pair of internal atomic states. This $3\times 3$ array can be understood as a determinant: each of the six terms appearing in the calculation of this determinant corresponds to one of the six types of triangles in Fig.~\ref{fig:Fig1}(d). The terms with positive (resp. negative) sign in the determinant calculation are for the upwards (resp. downwards) pointing triangles. 
}
\label{table}
\end{table}%

In order to characterize the possible non-trivial topology associated with the lattice in momentum space, we now evaluate the total phase  gained by an atom when it undergoes a series of Raman transitions $X\to Y\to Z \to X$ and performs a closed loop in momentum space. This corresponds to traveling around the three sides of one of the triangles of Fig.~\ref{fig:Fig1}(d). The resulting phase is different for upwards pointing triangles [such as the one of Fig.~\ref{fig:Fig1}(c)] and downwards pointing ones [like the triangles labelled $\alpha,\beta,\gamma$ in Fig.~\ref{fig:Fig1}(d)]. For an upwards pointing triangle, the global phase is always zero. Indeed moving around the sides of such a triangle involves absorption and stimulated emission of photons whose frequencies belong to the same triplet, {\it e.g.} the $\omega_i$ triplet for the triangle of Fig.~\ref{fig:Fig1}(c). Therefore each laser phase $\varphi_i$ (or $\varphi'_i$, $\varphi''_i$) enters both with a $+$ and a $-$ sign in the total accumulated phase around such a triangle, leading to a null result. 

Downwards pointing triangles on the other hand correspond to a non-trivial phase. Consider for example the clockwise oriented path around the sides of the triangle labelled $\alpha$ in Fig.~\ref{fig:Fig1}(d): (i)   The $X\to Y$ transition is accompanied by a change of atomic momentum $\bs q_2$, and it corresponds to a phase change $\varphi'_2-\varphi'_3$ (see Table \ref{table}); (ii) the $Y\to Z$ transition is along $\bs q_1$, with the phase change $\varphi''_1-\varphi''_2$; (iii) the $Z\to X$ transition is along $\bs q_3$, with the phase change $\varphi_3-\varphi_1$. As a result, the phase accumulated when traveling around the sides of triangle $\alpha$ is 
\begin{equation}
\Phi_\alpha=\varphi''_1-\varphi_1+\varphi'_2-\varphi''_2+\varphi_3-\varphi'_3.
\label{eq:triangle_phase}
\end{equation}
We can similarly calculate the phases $\Phi_{\beta,\gamma}$ for the two other downwards pointing triangles. Although the sum $\Phi_{\alpha}+\Phi_{\beta}+\Phi_{\gamma}$ is always zero, we can identify configurations such that each of these three phases takes a non-trivial value. For example the choice $\varphi_1=2\pi/3$, $\varphi_3=-2\pi/3$, and all other phases equal to zero yields
\begin{equation}
\Phi_{\alpha}=\Phi_{\beta}=\Phi_{\gamma}=2\pi/3 \quad \mbox{mod }2\pi.
\label{eq:triangle_phase_choice}
\end{equation} 
From now on, we will stick to this choice, together with the assumption that  $\Omega>0$, which is obtained for an alkali-metal atom by tuning the lasers between the $D_1$ and $D_2$ resonance lines.

The OFL formed in this way has a reciprocal lattice spanned by the
basis vectors ${\bm G}_1 = 3 {\bm q}_1$ and ${\bm G}_2 = {\bm
  q}_2$. The real space lattice vectors are ${\bm a}_1 =
\frac{2\pi}{3\sqrt{3}q}(\sqrt{3}\,\bs u_x + \bs u_y)$ and ${\bm a}_2 =
\frac{4\pi}{\sqrt{3}q}\bs u_y$, where $q = |{\bm q}_i| =
\sqrt{3}k$. This geometry is equivalent to that of the three-state
triangular flux lattice of Ref.~\onlinecite{coopermoessner}.  However,
the pattern of phases in the reciprocal space tight binding model
differs: here we have fluxes of $0$ and $2\pi/3$ in the upwards and 
downwards pointing triangles, as opposed to $\pi/3$ for
each \cite{coopermoessner}. Nevertheless, the physical properties of
the OFLs are very similar: in each unit cell of the real space lattice
the lowest energy dressed state experiences $N_\phi=1$ flux quantum;
the resulting bandstructure shows low energy bands that are analogous
to Landau levels.
In particular, the lowest energy band has Chern number of $1$, and
very narrow energy width, $W$, over a broad range of lattice depths
$\Omega$.  Here we focus on a lattice of depth $\Omega = 3E_{\rm R}$
[where $E_{\rm R} \equiv q^2/(2m)$ is the
recoil energy for atomic mass $m$]  close to which this bandwidth has a
(local) minimum of $W\simeq 0.015 E_{\rm R}$.  In view of this very
small bandwidth, the system is highly susceptible to the formation of
strongly correlated phases even for relatively weak interactions.

We have used exact diagonalization to study the groundstates of
interacting bosons occupying the lowest energy band of the OFL for
$\Omega = 3 E_{\rm R}$. (We neglect the population of higher bands,
since the gap to the next band is very large, $\Delta \simeq 46\, W$.)
We consider the bosons to interact via spin-independent contact
interactions, which is a good approximation for $^{87}$Rb. We
  write the two-dimensional coupling constant
 as $g_{\rm 2D} =
  \frac{\hbar^2}{m} {\tilde g}$, where 
${\tilde g}$ is  dimensionless.
  For atoms with 3D scattering length
$a_{\rm s}$ restricted to 2D by a harmonic confinement of oscillator
length $a_0$, and neglecting (sub)band mixing, this is ${\tilde
  g} = \sqrt{8\pi}a_{\rm s}/a_0$ \cite{hadzibabicdalibard}.
We study a finite system in a periodic geometry, with sides ${\bm
  L}_1 = N_1 {\bm a}_1$ and ${\bm L}_2 = N_2 {\bm a}_2$, where
$N_{1,2}$ are integers. The total flux is then $N_\phi =
N_1 N_2$, so for $N$ particles the Landau level filling factor is
$\nu\equiv N/N_\phi$.  The interacting many-particle states can be
classified by a conserved crystal momentum at $N_\phi$ points in the
Brillouin zone. We construct the Hamiltonian at each crystal momentum
and use a standard Lanczos method to determine the low energy
spectrum.

For very weak interactions, ${\tilde g}\ll 1$, the bosons form a
condensate in the minima of the band dispersion. However, our
numerical results show that, even for moderate interaction strength
${\tilde g}\gtrsim 0.2$, this (compressible) condensed phase is
replaced by strongly correlated (incompressible) FQH states at
  filling factors $\nu=1/2,2/3,3/4$ and 1.  Here, we focus on the FQH
  states at $\nu=1/2$ and $1$. (Results for $\nu=2/3, 3/4$ are
  described in the Supplementary Material.)

  Evidence for the appearance of incompressible phases is found by
  calculating the discontinuity in the chemical potential $\Delta \mu$
  for the groundstate: the difference between the chemical potential
  for adding a particle and that for removing a particle. A non-zero
  and positive $\Delta \mu$ indicates that the system is
  incompressible. To minimize finite-size effects we define\cite{cwg}
  $\Delta\mu \equiv N\left[E_{N+1}/(N+1)+E_{N-1}/(N-1) -
    2E_N/N\right]$, where $E_N$ is the groundstate energy for $N$
  particles.

\begin{figure}
\includegraphics[width=0.9\columnwidth]{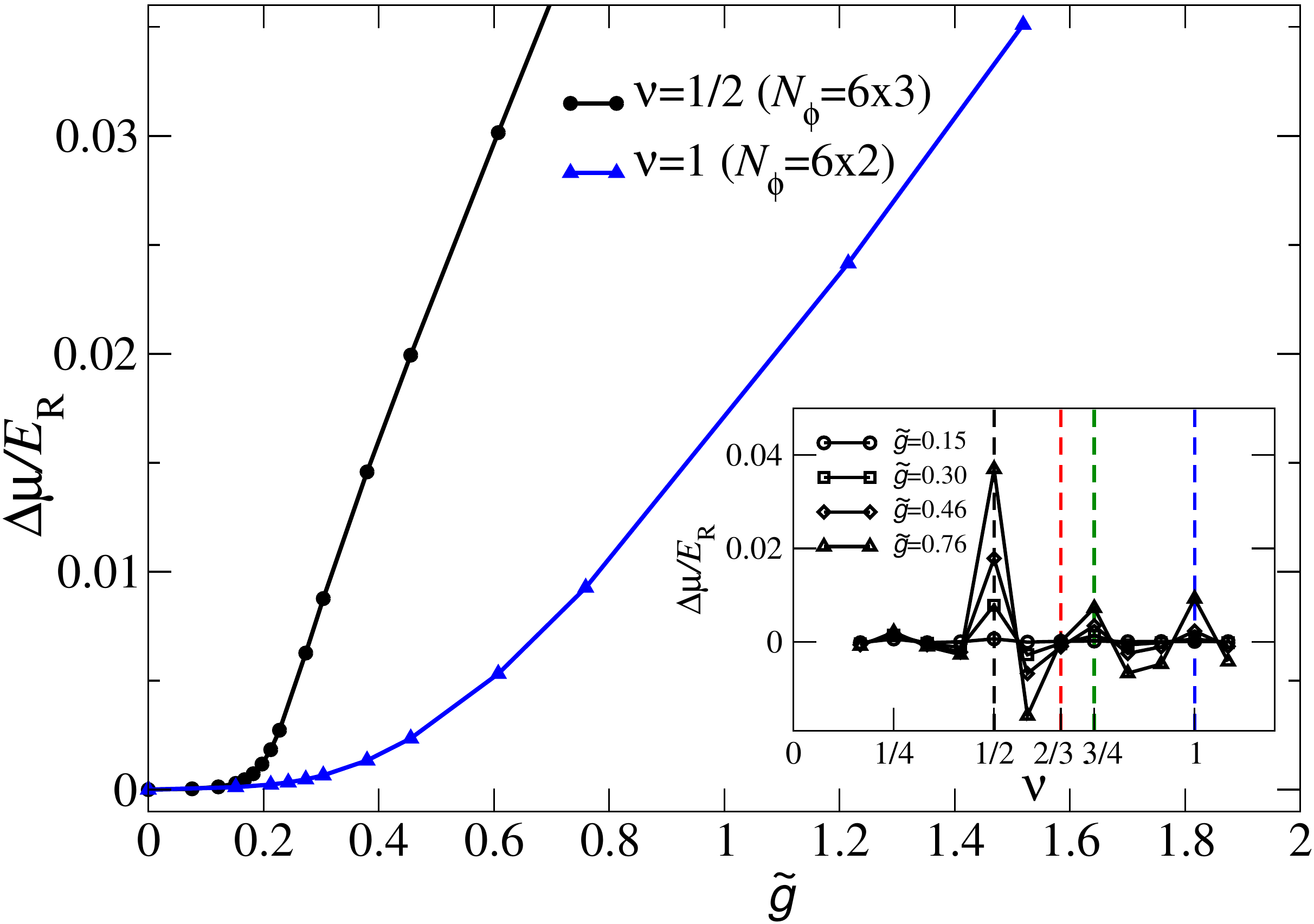}
\caption{\label{fig:incompressible}
  Incompressibility, as
  measured by the discontinuity in the chemical potential $\Delta \mu$
  defined in the text, as a function of interaction strength
${\tilde g}$ at several filling factors:
 $\nu=1/2$ (for $N=9$
bosons in a system of size 
$N_\phi = N_1 N_2 = 6\times 3$, circles);
and $\nu=1$ ($N=12$ in 
$N_\phi = 6\times 2$, triangles).
The inset shows $\Delta \mu$ as a function of filling
  factor for a series of interaction strengths ${\tilde g}$.}
\end{figure}

In Fig.~\ref{fig:incompressible} we plot the dependence of $\Delta
\mu$ on interaction strength ${\tilde g}$ at filling factors $\nu=1/2$
and $1$.  For $\nu=1/2$ there is an onset of incompressibility for
${\tilde g} \gtrsim 0.2$, and for $\nu=1$ incompressibility appears
for ${\tilde g}\gtrsim 0.4$. In the thermodynamic limit, $N\to
\infty$, the transitions from compressible $\Delta \mu=0$ to
incompressible $\Delta \mu>0$ should be sharp, and can even be
discontinuous for first-order transitions, but are rounded in Fig.~\ref{fig:incompressible} by finite-size effects.
The observed rises of $\Delta \mu$ are indications
of the approximate values of ${\tilde g}$ at which there are
transitions into the incompressible phases.  

\begin{figure}
\includegraphics[width=1.0\columnwidth]{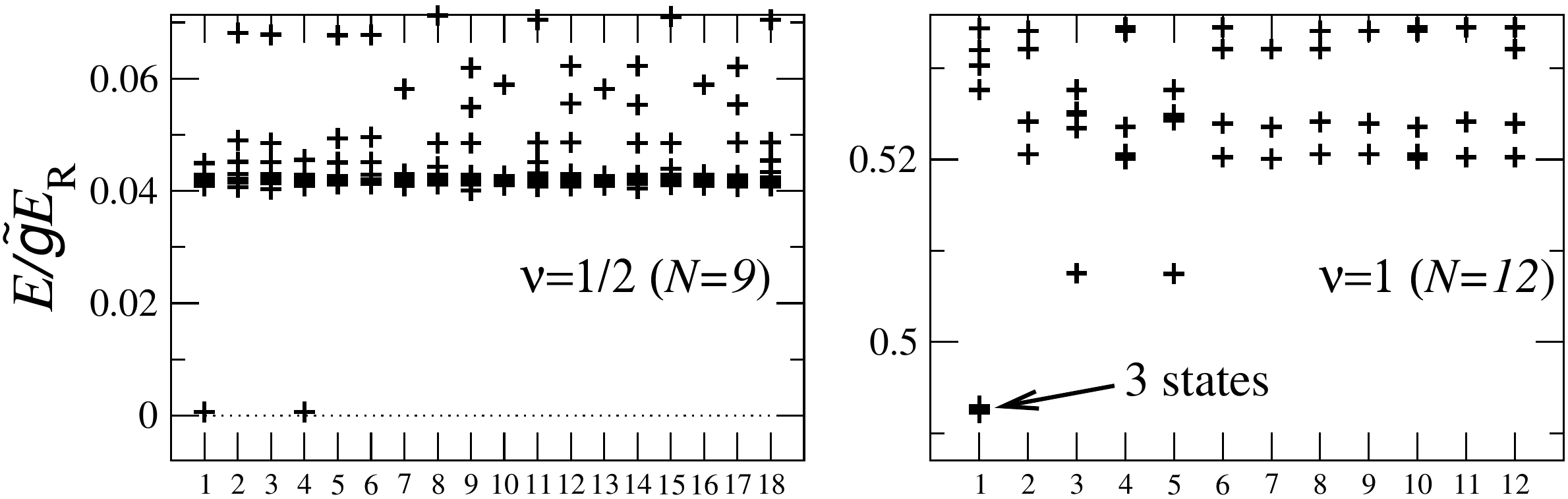}
\vskip-0.5cm
\hskip-3.5cm (a)\hskip4cm(b)
\caption{\label{fig:spectrum}
Low-energy spectra for 
the OFL with $\Omega/E_{\rm R} = 3$ in the strong-interaction limit ${\tilde g}\to \infty$ at filling factors 
(a) $\nu=1/2$ ($N=9$ bosons in $N_\phi = N_1N_2 = 6\times 3$)
and (b) $\nu=1$ ($N=12$ bosons in $N_1N_2 = 6\times 2$).
The crystal momentum ${\bm k} \equiv \alpha_1 {\bm G}_1/N_1 + \alpha_2 {\bm G}_2/N_2$ is
labelled by the 
index, $i=1+\alpha_1 + N_1 \alpha_2$ for $\alpha_1=0,\ldots,N_1-1$ and $\alpha_2=0,N_2-1$.
The quasi-degenerate groundstates have the expected multiplicities and
crystal momenta for the Laughlin state ($\nu=1/2$), and the Moore-Read state
($\nu =1$).}
\end{figure}

To explore the nature of these incompressible phases it is instructive
to study their (neutral) excitation spectra in the strong-interaction
limit ${\tilde g}\to \infty$.  These spectra, Fig.~\ref{fig:spectrum},
show all the expected properties of the bosonic Laughlin ($\nu=1/2$)
and Moore-Read  ($\nu=1$) states.  On this periodic geometry,
these topologically ordered incompressible phases should show
groundstate degeneracies (of 2 and 3 respectively) in the
thermodynamic limit, separated by an energy gap from the remaining
excitations.  As shown in Fig.~\ref{fig:spectrum}, even for these
finite systems these groundstate degeneracies appear clearly.  Results
on other system sizes and geometries (not shown) are consistent with
these results, confirming that these near degeneracies are robust
features, not imposed by symmetries, that characterize these topological phases.

The FQH states that we find for the OFL (at $\nu = 1/2,2/3,3/4,1$) are
the same as those found for contact interacting bosons in the
continuum lowest Landau level (LLL) \cite{advances}.  We have
established the equivalence of the phases of these two models by
studying the evolution of the many-body spectrum for a series of Bloch
wavefunctions that interpolate between those of the lowest band of the
OFL and those of the LLL. To do so, we consider a fictitious atom with
$N_{\rm s} =12$ internal states, and represent the LLL by the $N_{\rm
  s} = 12$ triangular OFL of Ref.~\protect\onlinecite{coopermoessner},
the lowest band of which has properties that are indistinguishable
from those of the LLL for suitable coupling $\Omega'\simeq 10 E_{\rm
  R}$ \cite{corrections}. We place 9 additional internal states at the
midpoints of the bonds of Fig.\ref{fig:Fig1}(d), coupled to each other
and to the original states $X,Y,Z$ by bonds of strength $\Omega'$ and
with $\pi/12$ flux through each new triangular plaquette \cite{flow}.
Choosing $(\Omega,\Omega') = (3(1-\lambda), 10\lambda)E_{\rm R}$ and
varying $\lambda$ leads to smooth interpolation of the lowest energy
band and the many-body spectrum, from those of the present model
($\lambda=0$) to those of the LLL ($\lambda=1$).  In all cases ($\nu =
1/2,2/3,3/4,1$) the energy gap remains open, showing that the phases
of these two models are the same \cite{scaffidimoller,wujainsun}.
Indeed, there is very little change in the spectrum, showing that this
OFL (with $\Omega = 3 E_{\rm R}$) is a very close representation of
the LLL. For example, for the LLL the $\nu=1/2$ state has zero
interaction energy, as the two-body correlation function vanishes
exactly at zero range. Here, the $\nu=1/2$ state in the OFL has
$E_N/N\simeq 7\times 10^{-5} {\tilde g} E_{\rm R}$ [see
Fig.~\ref{fig:spectrum}(a)] showing that the zero-range two-body
correlation function nearly vanishes.

To summarize we have proposed a robust atom-laser configuration that
can lead to FQH states of bosons in a well-accessible range of
parameters. The robustness of the setup is ensured by the absence of
need for a stabilization of the relative phases of different
beams \cite{Note2}.  The only phase difference to be controlled is
within each single-mode laser beam ($\varphi_i,
\varphi'_i,\varphi''_i$) and can be set by acoustic-optic modulators
driven by programmable function generators.  The lowest energy
  band is insensitive to fluctuations in the laser amplitudes around
  $\Omega = 3E_{\rm R}$, its bandwidth increasing by less than
  $10^{-3} E_{\rm R}$ within the range $\Omega/E_{\rm R} = 2 - 4$, and
  its topology remaining unchanged.  The minimal interaction strength
${\tilde g} \approx 0.2$ for obtaining FQH states corresponds to a 2D
confinement frequency of $\gtrsim 7\,\mbox{kHz}$ for Rb, which is
readily achieved in an optical lattice. A clear signal of the
formation of strongly correlated phases would be the appearance of
density plateaus (wedding cake structure) in in-situ images of the
gas, arising from incompressibility $\Delta \mu>0$.  This requires the
temperature to be smaller than $\Delta \mu$, which for the Laughlin state
we find  from Fig.~\ref{fig:incompressible} to be $\approx 0.02 E_{\rm R}$ for ${\tilde g} = 0.4$, that is
$10$\,nK for $^{87}$Rb.

\vskip0.1cm

\acknowledgments{  This work was supported by the Royal Society of  London, EPSRC Grant EP/J017639/1 (NRC), IFRAF and ANR (Grant AGAFON). We acknowledge useful discussions with J. Beugnon, L. Corman, C. Cohen-Tannoudji, F. Gerbier, G. M{\" o}ller, and S. Nascimb\`ene. }


\begin{thebibliography}{10}
\bibitem{external}
The three states $|X\rangle$, $|Y\rangle, |Z\rangle$ can also stand for different motional states, such as the first three vibrational states along the strongly confined direction $z$. However in this case, the hypothesis of a single, state-independent interaction strength should be relaxed.
\bibitem{corrections} Deviations from the continuum Landau level fall
  exponentially with $N_{\rm s}$ \cite{coopermoessner}, and are
  negligible for $N_{\rm s}=12$.
\bibitem{flow} This new model has net flux $\pi/3$ through each
    triangle of Fig.~\ref{fig:Fig1}(d). In connecting the new ($\Omega'$)
    bonds to those ($\Omega$) of the original model (with fluxes
    of $0$ and $2\pi/3$ through upwards and downwards pointing triangles),
    we assign flux $\pi/3$ through the loop formed by the two sets of bonds spanning each link ${\bm q}_3$.
\end{thebibliography}


\newpage
\null
\newpage

\setcounter{equation}{0}
\setcounter{figure}{0}

\renewcommand{\thefigure}{S\arabic{figure}}

\renewcommand{\theequation}{S\arabic{equation}}

\centerline{\textbf{Supplementary material}}
\vskip 5mm

\section{The basis set $|X\rangle,|Y\rangle,|Z\rangle$}
The components $\hat J_X$ and $\hat J_Y$ of a spin 1 angular momentum operator read in the standard eigenbasis  $\{|m_Z=+1\rangle, |m_Z=0\rangle, |m_Z=-1\rangle\}$  of $\hat J_Z$:
\begin{equation}
\hat J_X=\frac{1}{\sqrt 2}\begin{pmatrix}
0& 1 & 0\\
1 &0 &1\\
0 & 1 & 0
\end{pmatrix}
\quad
\hat J_Y=\frac{1}{\sqrt 2}\begin{pmatrix}
0& -i & 0\\
i &0 &-i\\
0 & i & 0
\end{pmatrix} .
\label{eq:JX_JY}
\end{equation}
In the standard eigenbasis $|m_Z=0,\pm 1\rangle$, we define the eigenvectors $|\alpha\rangle$ of $\hat J_\alpha$ ($\alpha=X,Y,Z$) with zero eigenvalue:
\begin{equation}
|X\rangle =\frac{1}{\sqrt 2}\begin{pmatrix}
-1 \\ 0 \\ 1
\end{pmatrix}
\quad
|Y\rangle =\frac{i}{\sqrt 2}\begin{pmatrix}
1 \\ 0 \\ 1
\end{pmatrix}
\quad
|Z\rangle =\begin{pmatrix}
0 \\ 1 \\ 0
\end{pmatrix}.
\label{eq:X_Y_Z}
\end{equation}
These three states also form an orthonormal basis set (denoted hereafter the Cartesian basis) and are such that 
\begin{equation}
\hat J_X|Y\rangle = i |Z\rangle, \quad \hat J_Y|Z\rangle = i |X\rangle, \quad \hat J_Z|X\rangle = i|Y\rangle .
\label{eq:circular}
\end{equation}
 For any real 3-vector $\bs u$, the state $u_X|X\rangle +u_Y|Y\rangle +u_Z|Z\rangle$ is the eigenstate of $\bs u\cdot \hat {\bs J}$ with eigenvalue 0. 

\section{Microwave dressing of a spin-1 ground state}

We consider an atom from the alkali-metal family with a nuclear spin $I=3/2$, so that its ground level is split in two hyperfine levels with angular momentum $F=1$ and $F=2$. The atom is irradiated with a linearly polarized microwave (mw), whose frequency is detuned by $\Delta_{\rm mw}$ from the frequency $\omega_{\rm hf}$ of  the $F=1 \leftrightarrow F=2$ hyperfine transition. The coupling between the mw and the atom (half-Rabi frequency) is defined by $\kappa_{\rm mw}=\mu_{\rm B} B_{\rm mw}/2$, where $\mu_{\rm B}$ is the Bohr magneton and $B_{\rm mw}$ is the amplitude of the oscillating mw magnetic field. We restrict to the case where $\kappa_{\rm mw} \ll \Delta_{\rm mw} \ll \omega_{\rm hf}$. We can then treat the atom-mw coupling using the rotating wave approximation and evaluate the shifts of the Zeeman states of the $F=1$ level using second-order perturbation theory. Taking the quantization axis  parallel to the polarization axis ($Z$) of the microwave, the shifts of the states $m_Z=0,\pm 1$ read \cite{Gerbier:2006}
\begin{equation}
\Delta E({m_Z})=\frac{\kappa_{\rm mw}^2}{\Delta_{\rm mw}} \left(1-\frac{m_Z^2}{4}\right).
\label{eq:mw_shift}
\end{equation}
The action of the mw on the $F=1$ hyperfine level can thus be described by the operator $\alpha +\beta \hat J_Z^2$, where $\beta=-\kappa_{\rm mw}^2/(4\Delta_{\rm mw})$.

Suppose now that a second mw, at a different detuning $\Delta'_{\rm mw}$ and with a linearly polarization along $X$,  also irradiates the atom. In the perturbative framework used above, the combined action of the two mws is described by the effective Hamiltonian $H_{\rm mw,eff}=\beta \hat J_Z^2 + \beta' J_X^2$, up to an additive constant. The eigenstates of  $H_{\rm mw,eff}$ are the states $|X\rangle, |Y\rangle,|Z\rangle$ with the energies $\beta$, $\beta+\beta'$, $\beta'$, respectively.

To estimate the required value for $B_{\rm mw}$, we start by noticing that when writing the atom-laser coupling [Eqn~(1) of the main text], we only take into account the resonant elements. For example we assume that the atom can undergo a transition $|X\rangle\to|Y\rangle$ via the absorption of a photon $\bs k_1$ and the emission of a photon $\bs k_2$, and we neglect the transitions $|X\rangle\to|Y\rangle$ occurring via the absorption of $\bs k_2$ and emission of  $\bs k_1$. This is legitimate when the two-photon coupling $\Omega$ is small compared to the energy difference between $|X\rangle$, $|Y\rangle$, $|Z\rangle$: $\Omega\ll \beta,\beta'$. We also note that the perturbative approach leading to (\ref{eq:mw_shift}) is valid only when $\kappa_{\rm mw}\ll \Delta_{\rm mw}$. Requiring that the two sides of each strong  
inequality differ by at least a factor 5, we find $\Omega \lesssim \kappa_{\rm mw}/100$. 
Consider as an illustration the case of Rubidium atoms for which the optimal two-photon coupling $\Omega=3 q^2/(2m)=9 k^2/(2m) \approx 2\pi \times 30$\,kHz. The microwave coupling $\kappa_{\rm mw}$ has to be $\gtrsim 3$\,MHz, corresponding to a microwave magnetic field of $\gtrsim 4$\,G. This is a large, but still realistic value, especially if one uses small resonant loops to increase the value of  $B_{\rm mw}$ at the location of the atomic sample.

\section{The light-shift operator for an alkali-metal atomic scheme}

\begin{figure}[t]
\begin{center}
\vskip 5mm
\includegraphics[height=2.5cm]{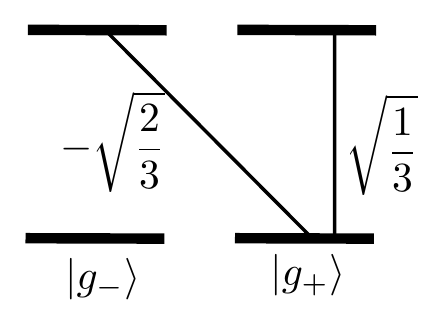}
\vskip 5mm
\includegraphics[height=2.5cm]{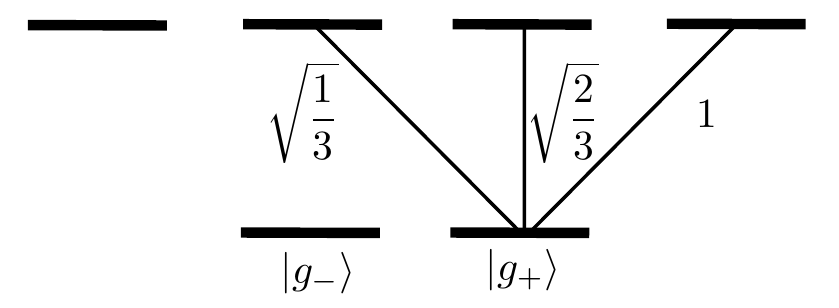}
\caption{Amplitude of the couplings for the two components D$_1$ (top) and D$_2$ (bottom) of the resonance line of an alkali atom. Here the spin of the nucleus is supposed to be zero. The coupling amplitudes for the $g_-$ state take symmetric values with respect to those indicated for the $g_+$ state.}
\label{Fig:CG}
\end{center}
\end{figure}

We consider an atom that is irradiated by a monochromatic laser beam of frequency $\omega$. The light-shift operator $\hat{V}$ gives the restriction of the atom-laser coupling to the ground atomic level $g$ at first order in laser intensity \cite{cct:1972}. In the absence of electron and nuclear spins, the ground and excited states that are involved in the resonant transition of an alkali-metal atom have an angular momentum $J_g=0$ and $J_e=1$, respectively. The atom-laser coupling can be written
\begin{eqnarray}
\hat{U} &=& \sum_{\alpha=X,Y,Z} \kappa_\alpha |e, \alpha \rangle \langle g| + \mbox{H.c.}\\
&=& \sum_{m=0,\pm 1} \kappa_m |e, m \rangle \langle g| + \mbox{H.c.}
\label{}
\end{eqnarray}
Here the $\kappa_\alpha$, $\alpha=X,Y,Z$, (resp. $\kappa_m$, $m=0,\pm 1$) denote the  atom-laser coupling strengths (half-Rabi frequencies) in the Cartesian (resp. standard) basis with
\begin{equation}
\kappa_\pm=\frac{1}{\sqrt 2}(\mp \kappa_X + i\kappa_Y),\qquad \kappa_0=\kappa_Z.
\label{}
\end{equation}
In this ``no-spin" case, the light-shift operator simply describes the displacement of the ground state by the quantity $|\bs \kappa|^2/\Delta$, where $\Delta$ is the detuning of the laser with respect to the atomic transition, $\bs \kappa=(\kappa_X,\kappa_Y,\kappa_Z)$  and
\begin{equation}
|\bs \kappa|^2= \sum_{\alpha=X,Y,Z} |\kappa_\alpha|^2=\sum_{m=0,\pm 1} |\kappa_m|^2 .
\label{}
\end{equation}

We now take the spin of the valence electron into account, but still assume that the nucleus has spin zero. The ground level is then a two-dimensional manifold (angular momentum $J_g=1/2$). Because of the fine-structure Hamiltonian, the resonance line of the atom is split into two components D$_1$ and D$_2$, corresponding to the transition from the ground state to the excited states with total angular momentum $J_e=1/2$ and $J_e=3/2$, respectively. 

Let us focus first on the light-shift operator associated with the D$_1$ transition,  which is dominant if the detuning $\Delta_1=\omega-\omega_1$ of the laser excitation from the D$_1$ line (frequency $\omega_1$) is much smaller than the detuning $\Delta_2$ from the D$_2$ line. Using the well-known Clebsch-Gordan coefficients (see {\it e.g.} Figure~\ref{Fig:CG})  and taking $Z$  as quantization axis, we get in the corresponding $\{ |g_+\rangle, |g_-\rangle  \}$ basis:
\begin{equation}
\hat V= \frac{1}{3\Delta_1}
\begin{pmatrix}
|\kappa_0|^2+2|\kappa_-|^2 & -\sqrt 2(\kappa_0\kappa_-^* + \kappa_+ \kappa_0^*)\\
 -\sqrt 2(\kappa_-\kappa_0^* + \kappa_0 \kappa_-^* ) &
 |\kappa_0|^2+2|\kappa_+|^2
\end{pmatrix} .
\label{eq:first_express_U}
\end{equation} 
This coupling can be written in a compact form
\begin{equation}
\hat V= \frac{|\bs \kappa|^2}{3\Delta_1}  \hat 1+ \bs B_1\cdot \hat{\bs S},
\label{eq:second_express_U}
\end{equation}
where $\hat 1$ and $\hat S$ are the identity and spin operators for the spin 1/2 ground state, respectively, and where the (real) effective field $\bs B_1$ for the D$_1$ transition is:
\begin{equation}
\bs B_1=- \frac{2i}{3\Delta_1} \bs \kappa \times \bs \kappa^*.
\label{eq:effective_field}
\end{equation}

We now take into account both the D$_1$ and D$_2$ transitions. A straightforward generalization of the preceding calculation leads to
\begin{equation}
\hat V=  A\, \hat 1+ \bs B\cdot \hat{\bs S},
\label{eq:third_express_U}
\end{equation}
with 
\begin{equation}
A=\frac{|\bs \kappa|^2}{3} \left( \frac{1}{\Delta_1}+\frac{2}{\Delta_2} \right)
\label{eq:scalar}
\end{equation}
and
\begin{equation}
\bs B= \frac{2i}{3} \left( \frac{1}{\Delta_2}- \frac{1}{\Delta_1}\right) \bs \kappa \times \bs \kappa^*.
\label{eq:effectiveB}
\end{equation}

As a final step we take into account the nucleus spin $I$. As above we consider the case $I=3/2$, which leads to a splitting of the ground level into two hyperfine states of angular momentum $F=1$ and $F=2$. Here we consider the $F=1$ state, with the three Zeeman sublevels
\begin{eqnarray*}
|F=1, m_F=\pm 1\rangle &=& \mp \frac{\sqrt 3}{2}|\mp \frac{1}{2};\pm\frac{3}{2}\rangle \pm \frac{1}{2}|\pm \frac{1}{2};\pm\frac{1}{2}\rangle, \\
|F=1, m_F=0\rangle &=& -\frac{1}{\sqrt 2} |-\frac{1}{2};\frac{1}{2}\rangle +\frac{1}{\sqrt 2}|\frac{1}{2};-\frac{1}{2}\rangle,
\end{eqnarray*}
where the state $|m_e;m_n\rangle$ is labelled by the quantum numbers for the projection along the $Z$ axis of the electron spin ($m_e$) and nuclear spin ($m_n$). We suppose that the detunings $\Delta_1$ and $\Delta _2$ of the laser with respect to the excited states $J_e=1/2$ and $J_e=3/2$ are large compared to the hyperfine splittings of these levels. The calculation of the restriction of the coupling (\ref{eq:third_express_U}) to the $F=1$ ground level then gives:
\begin{equation}
\hat V= A\,  \hat 1 +  \bs B'\cdot \hat{\bs F},
\label{eq:fourth_express_U}
\end{equation}
where 
\begin{equation}
\bs B'=-\frac{1}{4}\bs B=\frac{i}{6} \left( \frac{1}{\Delta_1}- \frac{1}{\Delta_2}\right) \bs \kappa \times \bs \kappa^*.
\label{eq:effectiveBfinal}
\end{equation}
Using (\ref{eq:circular}) the vector part of the coupling $\hat U_{\rm vec}=\hat U-A\,\hat 1$ can be written in the Cartesian basis for the $F=1$ ground state:
\begin{equation}
\hat  V_{\rm vec} = i \left( B'_Z |Y\rangle \langle X| + B'_X |Z\rangle \langle Y| + B'_Y |X\rangle \langle Z|\right) +\mbox{H.c.}
\label{eq:expanded_coupling}
\end{equation}

\begin{figure}[t]
\begin{center}
\vskip 5mm
\includegraphics[width=4cm]{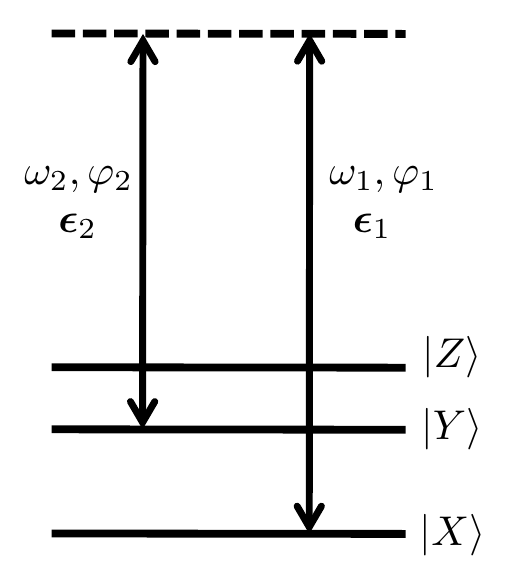}
\caption{Laser scheme providing a resonant Raman coupling between two sublevels of the $F=1$ ground state.}
\label{fig:laser_scheme}
\end{center}
\end{figure}

\section{Hamiltonian for resonant Raman transitions}

Consider now a scheme such as the one of Fig.~\ref{fig:laser_scheme}, where an external field lifts the degeneracy between the three states of the Cartesian basis, and where two monochromatic light waves with the same amplitude (characterized by $\kappa>0$) and with frequency, phase, wave vector and polarizations $\omega_i,\varphi_i,\bs k_i, \bs \epsilon_i$, $i=1,2$ induce a resonant Raman coupling between $|X\rangle$ and $|Y\rangle$. 

The transition from $|X\rangle$ to $|Y\rangle$ occurs resonantly via the absorption of a photon in wave 1 and the stimulated emission of a photon in wave 2. Keeping only this resonant process, the relevant contribution to the cross-product  $\bs \kappa \times \bs \kappa^*$ entering in (\ref{eq:effectiveBfinal}) is $\kappa^2 \left(\bs \epsilon_1 e^{i(\bs k_1\cdot\bs r+\varphi_1)}\right)\times \left(\bs \epsilon_2 e^{i(\bs k_2\cdot \bs r +\varphi_2)} \right)^*$, where $\bs r$ is the position of the atom.

In this work we restrict to waves with linear (real) polarizations in the $xy$ plane, making an angle of $2\pi/3$ with each other. In this case $\bs \epsilon_1\times \bs \epsilon_2=(\sqrt{3}/2)\bs u_z$. The projection of $\bs u_z$ on each basis vector of the trihedron $X,Y,Z$ is $1/\sqrt 3$, and the value of $B'_Z$ that is relevant for the resonant coupling between  $|X\rangle$ and $|Y\rangle$ (see (\ref{eq:expanded_coupling})) is thus
\begin{equation}
B'_Z=\frac{i \kappa^2}{12} \left( \frac{1}{\Delta_1}- \frac{1}{\Delta_2}\right) e^{i(\bs q_1\cdot\bs r+\varphi_1-\varphi_2)},
\label{}
\end{equation}
with $\bs q_1=\bs k_1-\bs k_2$. 
This leads to the resonant part of the vector light-shift operator:
\begin{equation}
\hat V_{\rm vec}^{\rm (res)}= - \Omega\, e^{i(\bs q_1\cdot\bs r+\varphi_1-\varphi_2)}\, |Y\rangle \langle X| +\mbox{H.c.},
\label{}
\end{equation} 
where we have set
\begin{equation}
\Omega= \frac{\kappa^2}{12} \left( \frac{1}{\Delta_1}- \frac{1}{\Delta_2}\right) .
\label{}
\end{equation}
The coupling strength $\Omega$ is positive when the laser is tuned between the D$_1$ and the D$_2$ lines ($\Delta_1>0>\Delta_2$), and negative otherwise.

Consider for example the case of Rubidium atoms and choose the
detuning such that the scalar part of the atom laser coupling
(\ref{eq:scalar}) vanishes: $\Delta_2=-2\Delta_1$, corresponding to
the wavelength $\lambda=790$\,nm ($\lambda_1=795$\,nm,
$\lambda_2=780$\,nm). The optimal two-photon coupling is $\Omega=3
E_{\rm R}=9 k^2/(2m) \approx 2\pi \times 30$\,kHz and the photon
scattering rate is $\gamma\approx 9\times \Gamma
\kappa^2/(2\Delta_1^2) \approx 20$\,s$^{-1}$, where the factor 9
accounts for the number of monochromatic beams shining on the
atoms. The time scale for establishing a many-body state such as the
Laughlin state can be estimated as the inverse of the corresponding
gap $\Delta\mu^{-1}$. Taking $\Delta \mu=0.02\,E_{\rm R}$ as a typical
value (see Fig.~2 of the main text), we obtain
$\Delta\mu^{-1}\approx 1\,$ms.  The heating due to photon scattering
should then play a minor role during this time.

\section{Additional Numerical Results}

We present in this section some additional numerical results alluded to in the main text.

\begin{figure}
\includegraphics[width=0.9\columnwidth]{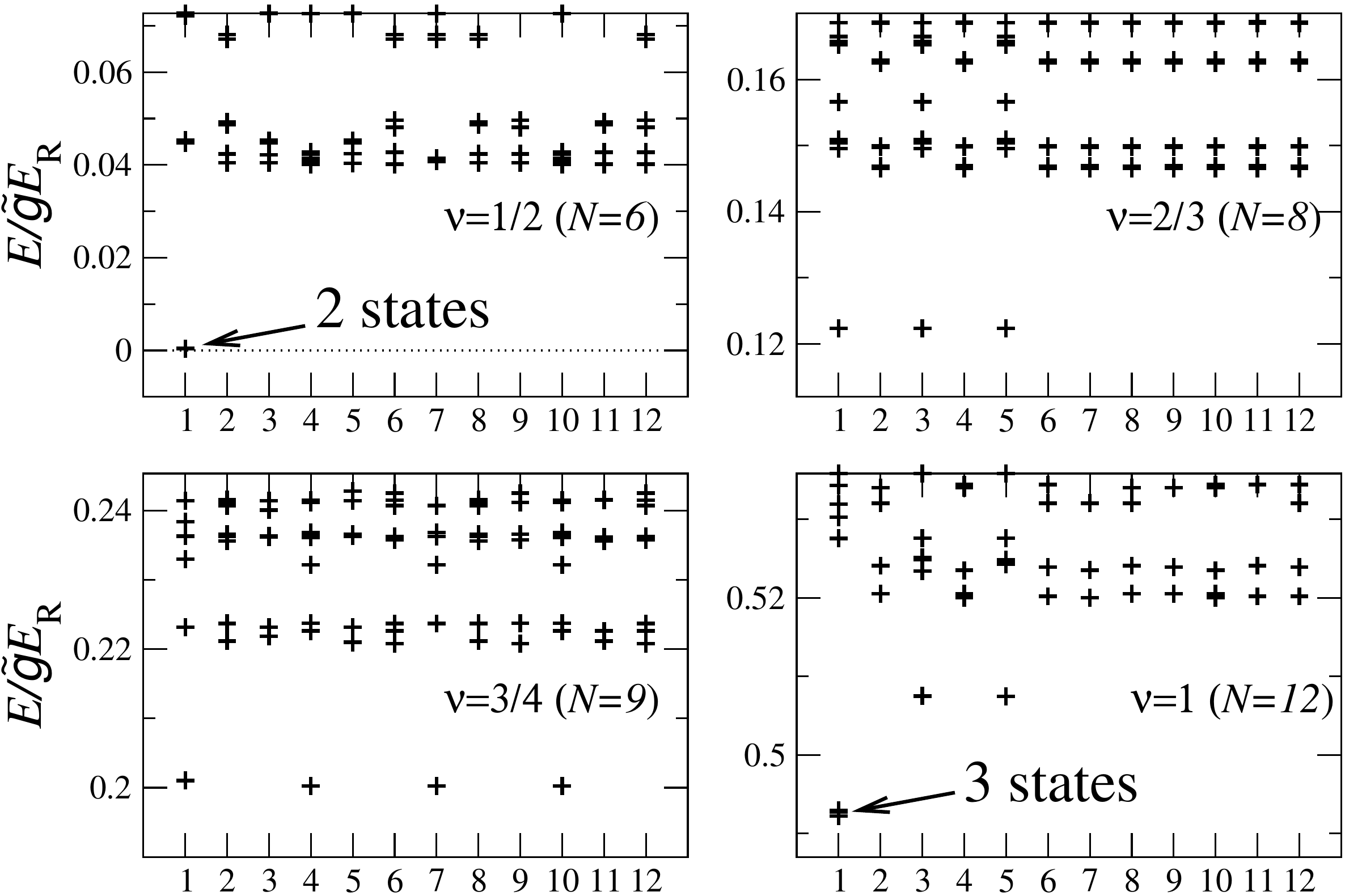}
\caption{\label{fig:spectrumall} Low-energy spectra for the OFL with $\Omega/E_{\rm R} = 3$ in the strong-interaction limit
  ${\tilde g}\to \infty$ for a system of size $N_\phi = N_1N_2 =
  6\times 2$ at filling factors for which the groundstate is
  incompressible.  The crystal momentum ${\bm k} \equiv \alpha_1 {\bm
    G}_1/6 + \alpha_2 {\bm G}_2/2$ is labelled by the index,
  $i=1+\alpha_1 + 6 \alpha_2$ for $\alpha_1=0,\ldots,5$ and
  $\alpha_2=0,1$.  The quasi-degenerate groundstates have the expected
  multiplicities and crystal momenta for the Laughlin state
  ($\nu=1/2$), composite fermion states ($\nu=2/3$, $3/4$), and the
  Moore-Read state ($\nu =1$).}
\end{figure}

In Fig.~\ref{fig:spectrumall}, we present the excitation spectra for a
system of size $N_1\times N_2 = 6\times 2$ at all filling factors
($\nu=1/2,2/3,3/4$ and $1$) for which we find evidence for
incompressible FQH states. These show the expected features of the
Laughlin state ($\nu=1/2$), composite fermion states ($\nu = 2/3,3/4$)
and Moore-Read state ($\nu=1$) of bosons.  On this periodic
geometry, these topologically ordered incompressible phases should
show groundstate degeneracies (of 2,3,4 and 3 respectively) in the
thermodynamic limit, separated by an energy gap from the remaining
excitations. This is indeed the case, as may be seen clearly in
  Fig.~\ref{fig:spectrumall}.

\begin{figure}
\includegraphics[width=0.9\columnwidth]{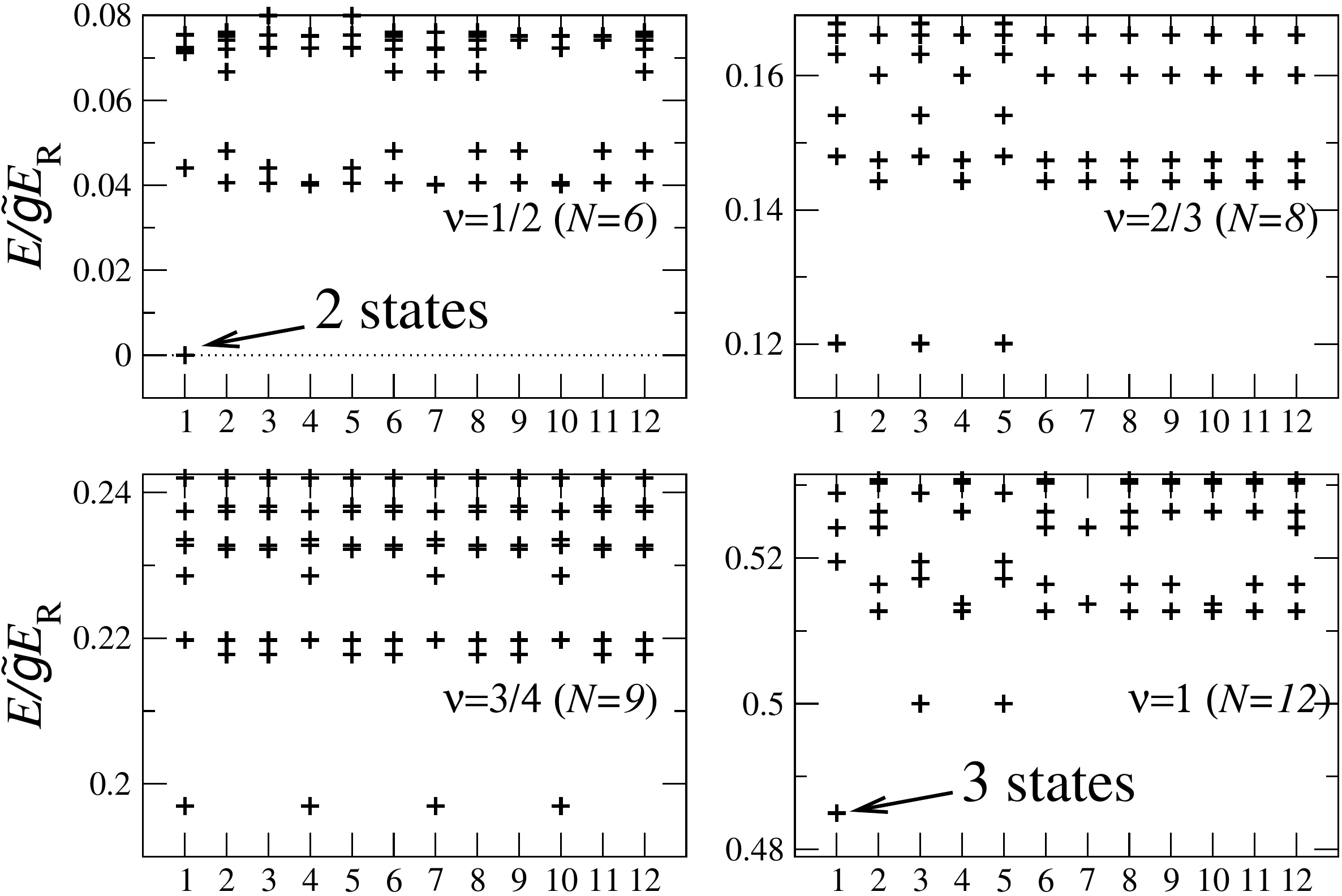}
\caption{\label{fig:lll} Low-energy spectra for contact-interacting
  bosons in the LLL for a system of the same size and
  geometry as Fig.~\protect\ref{fig:spectrumall} ($N_\phi=12$), at
  filling factors $\nu=1/2,2/3,3/4,1$ for which the groundstate is
  incompressible.  The crystal momentum is labelled as for
  Fig.~\protect\ref{fig:spectrumall}.}
\end{figure}

In Fig.~\ref{fig:lll}, we present the equivalent excitation spectra
for a lowest band formed from states in the lowest Landau level
(LLL). As described in the main text, we represent the LLL by the
$N_{\rm s}=12$ OFL lattice of Ref.~\onlinecite{coopermoessner2}, the lowest
band of which has properties that are indistinguishable from those of
the LLL for a lattice coupling of $\Omega' = 10 E_{\rm R}$. This very
close equivalence arises from the very rapid convergence of the
properties of the $N_{\rm s}$-state OFL lattice of
Ref.~\onlinecite{coopermoessner2} to those of a charged particle in a
uniform magnetic field, the spatial fluctuations of the energy and
effective magnetic field experienced by the lowest energy dressed
state of the OFL falling exponentially with increasing $N_{\rm s}$, and
already negligible for $N_{\rm s}=12$ \cite{coopermoessner2}.  Interpolation
between the $N_{\rm s}=3$ OFL of this paper and the LLL leads to a smooth
evolution both of the single-particle levels of the lowest bands, and
of the many body spectrum for bosons occupying this lowest band. This
establishes that the FQH phases of these two systems are the same.
Moreover, the spectra of the OFL (Fig.~\ref{fig:spectrumall}) differ
only very slightly from those of the LLL (Fig.~\ref{fig:lll}).  The
main qualitative difference is that some approximate degeneracies in
Fig.~\ref{fig:spectrumall} become exact degeneracies connected to the
many-body translational symmetry of the LLL \cite{haldanemtm}.
Furthermore, the approximate 3-fold groundstate degeneracy of
Fig.~\ref{fig:spectrumall}(d) becomes an exact degeneracy in the LLL
Fig.~\ref{fig:lll}(d).  This is due to a $\pi/3$ rotational symmetry
of the LLL in this geometry of $N_1\times N_2 = 6\times 2$ (for which
the sides of the simulation cell have equal length $|{\bm L}_1| =
|{\bm L}_2|$), which transforms the three groundstates of the
Moore-Read state in the LLL 
\begin{center}
  \includegraphics[width=0.7\columnwidth]{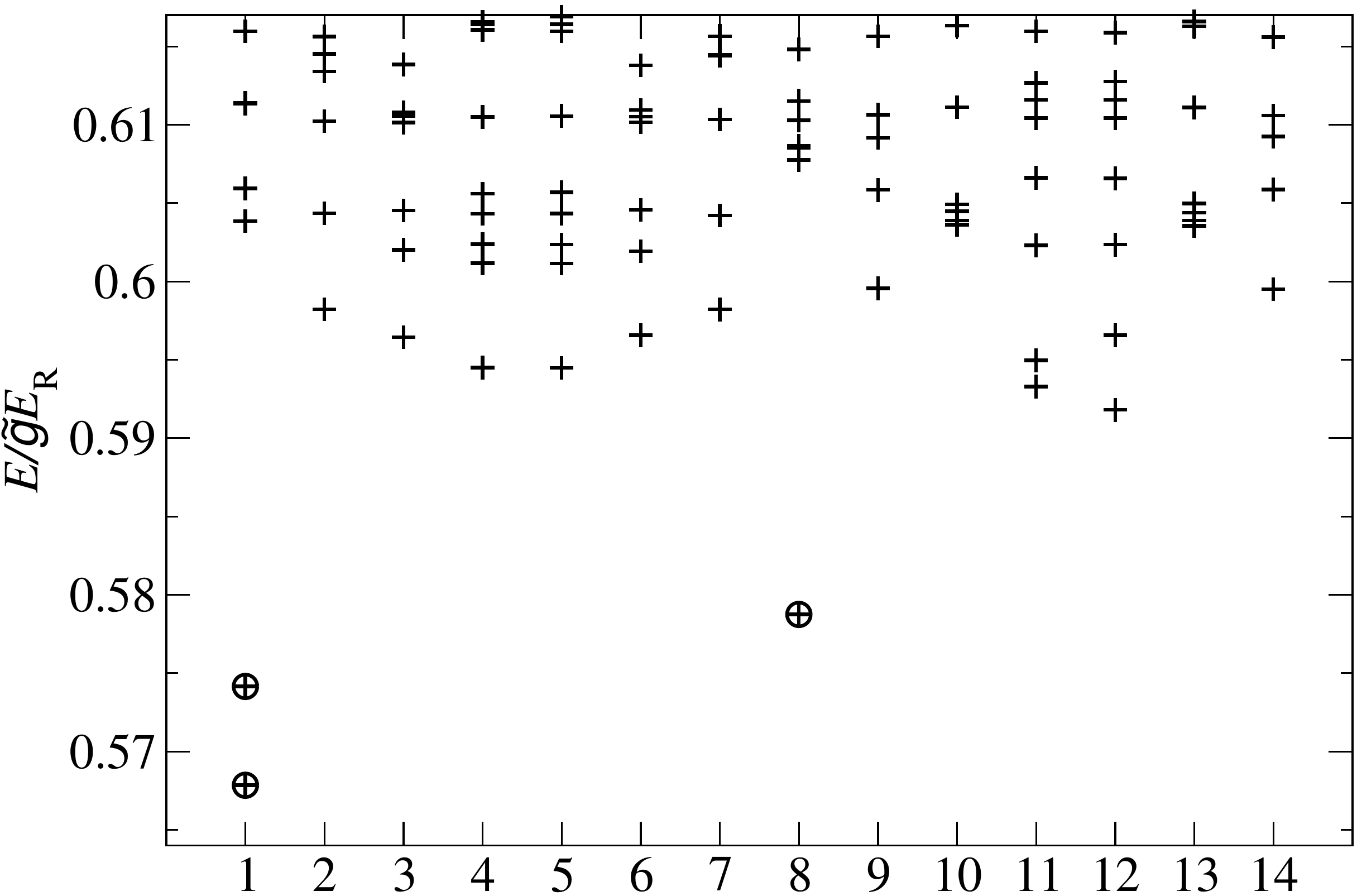}
  \captionof{figure}{\label{fig:nu1again} Low-energy spectrum for the OFL
with $\Omega/E_{\rm R} = 3$ in the strong-interaction
    limit ${\tilde g}\to \infty$ for a system of size $N_\phi = N_1N_2
    = 7\times 2$ at filling factor $\nu=1$.  The crystal momentum
    ${\bm k} \equiv \alpha_1 {\bm G}_1/7 + \alpha_2 {\bm G}_2/2$ is
    labelled by the index, $i=1+\alpha_1 + 7 \alpha_2$ for
    $\alpha_1=0,\ldots,6$ and $\alpha_2=0,1$.  The three
    quasi-degenerate groundstates (marked by circles) are at the
    expected crystal momenta for the Moore-Read state. No symmetry
    protects these quasi-degenerate levels. The same quasi-degeneracy
    appears for contact interacting bosons in the 
    LLL \protect\cite{advances2}.}
\end{center}
into each other. We emphasize that, in
general, this three-fold quasi-degeneracy is unrelated to any symmetry, but is an
emergent quasi-degeneracy in the thermodynamic limit. This is
evidenced by studies on other system sizes and geometries, just as in
the LLL \cite{advances2}. For example, Fig.~\ref{fig:nu1again} shows
the spectrum for the OFL for $N=14$ particles in a system of size
$N_1\times N_2 = 7\times 2$ for which no symmetry relates the three
quasi-degenerate groundstates.

\end{document}